\documentclass[usenatbib,usegraphicx,useAMS]{mn2e}
\usepackage{times}

\newcommand{\lsim}{\mbox{$_<\atop^{\sim}$}}

\newcommand{\kp}{$K^\prime$}
\newcommand{\kpd}{$K^\prime$-}
\newcommand{\mum}{$\,\mu$m}
\newcommand{\arcm}{$^{\prime}$}
\newcommand{\arcs}{$^{\prime\prime}$}

\newcommand{\ms}{MS\,0451.6$-$0305}
\newcommand{\chandra}{\textsl{Chandra}}
\newcommand{\spitzer}{\textsl{Spitzer}}
\newcommand{\hst}{\textsl{HST}}

\begin{document}

\title[A gravitationally lensed sub-millimetre arc in MS0451.6$-$0305]
{The nature of a gravitationally lensed sub-millimetre arc in  MS0451.6$-$0305:
two interacting galaxies at $z\sim 2.9$?}

\author[Borys et al.]{
\parbox[t]{\textwidth}{
\vspace{-1.0cm}
Colin Borys$^{1}$,
Scott Chapman$^{1}$,
Megan Donahue$^{2}$,
Greg Fahlman$^{3}$,
Mark Halpern$^{4}$,
Jean-Paul Kneib$^{1,5}$,
Peter Newbury$^{4}$,
Douglas Scott$^{4}$,
Graham P. Smith$^{1}$
}
\vspace*{6pt}\\
$^{1}$ California Institute of Technology, Pasadena, CA 91125, USA\\
$^{2}$ Michigan State University, East Lansing, MI 48824, USA\\
$^{3}$ Herzberg Institute of Astrophysics, Victoria, BC, Canada\\
$^{4}$ Department of Physics \& Astronomy, University of British  
Columbia,Vancouver, BC, Canada \\
$^{5}$ Observatoire Midi-Pyr\'en\'ees, 14 Avenue E., Belin, 31400,  
Toulouse, France
\vspace*{-0.5cm}}

\date{Submitted 10 Dec 2002}

\maketitle

\begin{abstract}
We present a new SCUBA image of the cluster \ms, which exhibits
strong, extended sub-mm flux at 850\mum.  The most striking feature
in the map is an elongated region of bright sub-mm emission, with a flux
density of ${\sim}\,10\,$mJy over several beam-sizes.  This region is
apparently coincident with a previously known optical arc (which turns
out to be a strongly lensed Lyman Break Galaxy at $z=2.911$), as well as
with a newly identified multiply imaged ERO (Extremely Red Object) pair 
predicted to be at a similar, if not identical redshift.  
By combing a detailed lensing model
with deep images from \hst, \chandra, CFHT, JCMT, and spectra from the
VLT, we conclude that both the strongly lensed optical arc and ERO
systems have properties consistent with known sub-mm
emitters.  Using a simple model for the two sources,
we estimate that the multiply lensed EROs contribute
the majority of the flux in the SCUBA lensed arc. Correcting for the lensing
amplification, we estimate that the inherent 850\mum\ fluxes for both
objects are \lsim0.4\,mJy.  
If the LBG and ERO pair are truly at the same redshift, then they
are separated by only $\sim10\,$kpc in the source plane, and hence
constitute an interacting system at $z\sim2.9$.
Higher angular resolution observations
in sub-mm/mm will permit us to more accurately separate the 
contribution from each candidate, and better understand the 
nature of this system.
\end{abstract}

\begin{keywords}
gravitational lensing --
galaxies: clusters : individual : \ms\ --
submillimetre --
methods: numerical --
techniques: image processing
\vspace*{-1.25cm}
\end{keywords}

\section{Introduction}
\label{sec:intro}
Since the installation of the Submillimetre Common User Bolometer
Array (SCUBA; Holland et al.~1999) at the JCMT telescope, hundreds of
luminous dusty galaxies have been detected (see Blain et al.~2002 for
a review). Such observations are still far from routine, however, with
the detection rate in random fields being typically one or two sources
of modest signal-to-noise ratio (SNR) per shift of telescope time.
Bright sources, which are easier to obtain detailed sub-mm/mm follow-up
observations for using heterodyne receivers or bolometers tuned to
other wavelengths, are correspondingly rarer.  This, coupled with the
fact that the dust responsible for the sub-mm luminosity absorbs
radiation at other wavelengths, means that SCUBA detected galaxies are
often extremely faint in the optical.  Hence, redshifts, morphologies
and spectral energy distributions (SEDs) have proven elusive for most
sources.

One technique which can help, pioneered by Smail, Ivison \& Blain
(1997), is to use the lensing amplification of rich galaxy clusters to
boost the detectability of background sub-mm galaxies. The rate at
which sources are detected is increased by a factor ${\sim}\,3$, with
an improved prospect of discovering the occasional bright object at
relatively high SNR.  Here we present detailed observations of just
such a source in the field of the massive cluster \ms, which we
discovered in an earlier survey of 9 rich cluster fields (Chapman et
al.~2002a, hereafter C02).  This paper uses new SCUBA data, as well as
rigorous chop deconvolution, in order to produce a map free from the
chopping artifacts common in sub-mm observations.

The sub-mm emission in this image is extended along a giant optical
arc discovered in this cluster by Luppino et al.~(1999, hereafter
L99), and thus there is little doubt that
the SCUBA emission is lensed as well.  In this paper, we combine deep
optical, near-IR (NIR), and X-ray data with a detailed lensing model
in order to interpret the sub-mm emission in this remarkable SCUBA
field.

\section{Observations}
\ms\ is the most luminous X-ray cluster in the Einstein Observatory
Extended Medium Sensitivity Survey (EMSS; Gioia et al.~1990).  It is a
$z=0.55$ cluster, which has
been studied extensively in the optical (Gioia \& Luppino~1994,
hereafter L99) and X-Ray (Molnar et al.~2002, Donahue et al.~2003;
hereafter D03).  It makes an ideal SCUBA target because it is
relatively compact compared to the SCUBA array size, and the strong
radio Sunyaev-Zel'dovich (SZ) decrement (Reese et al.~2000) suggests
that the SZ \textsl{increment} at 850\mum\ might be detectable. 
The SZ and X-ray signals clearly indicate that this is a massive
cluster, and the presence of a giant optically detected arc 
unambiguously verifies that it is lensing background galaxies.
Thus SCUBA observations have the potential to detect a
combination of diffuse SZ emission, background lensed sources, and
galaxies in the cluster itself.

\subsection{Sub-mm observations}
SCUBA `jiggle-mode' maps of the region were taken simultaneously at
850\mum\ and 450\mum\ in two separate JCMT runs; The first was
on 03 September 1998 and the second was 10-11 November 1999.
Each has approximately the same observing time of ${\simeq}\,22\,$ksec, 
but the weather in 1999 was significantly better, with an average
atmospheric opacity of $\tau_{850}\simeq 0.25$, compared to 0.40 in
1998.  In addition, upgrades after the 1998 run improved SCUBA's
overall sensitivity by roughly 10\% at 850\mum\ and 50\% at 450\mum.

To remove the strong atmospheric emission component of the data, the
JCMT secondary mirror `chops' between the target and two nearby
reference positions on the sky.  Since the sky emission is common to
all positions, but the astronomical target is not, the difference
between the data at these positions removes the majority of the common
mode noise. Three different chop directions, all using 30 arcsecond
throws, were used in 1998. Analysis of these data, revealed that the
region was excessively rich in bright sub-mm sources, raising concern
that the chopping may have removed some of the measured signal.
The sub-mm map of \ms\ presented in C02 use these data exclusively.

In an effort to detect the Sunyaev-Zel'dovich (SZ) increment in the
cluster, the 1999 observations were taken with
very wide chop throws (180 arcsecond), so that signal within the 2
arcminute SCUBA field of view would not be chopped out.  Note that
chopping onto sources {\it outside\/} the field is unlikely to be a
concern, since the chop was held at constant azimuth and therefore sky
rotation would quickly move sources out of the chopped position.
Zemcov et al.~2003 use the 1999 data only, since the data in C02
used as a chop throw too small to be useful for SZ studies.  By comparing
the maps made from each set, it is clear that 1998 data has, in fact, 
chopped out some signal inside the field.

The map presented here combines these two datasets, and given the 
differences between the 1998 and 1999 maps, it seems 
appropriate to be fairly comprehensive in describing the
map-making procedure which we adopted. This `direct-inversion' method
is described in the appendix.  It could be adapted for all SCUBA
jiggle-maps which have in-field chopping.  However, it is much more
computationally intensive than standard approaches, and will only lead
to mild improvements for general `blank' fields.  On the other hand,
for fields with multiple or extended sources, a method of this sort is
essential in order to recover the correct fluxes across the map.

\begin{figure*}
\includegraphics[width=0.9\textwidth,angle=0]{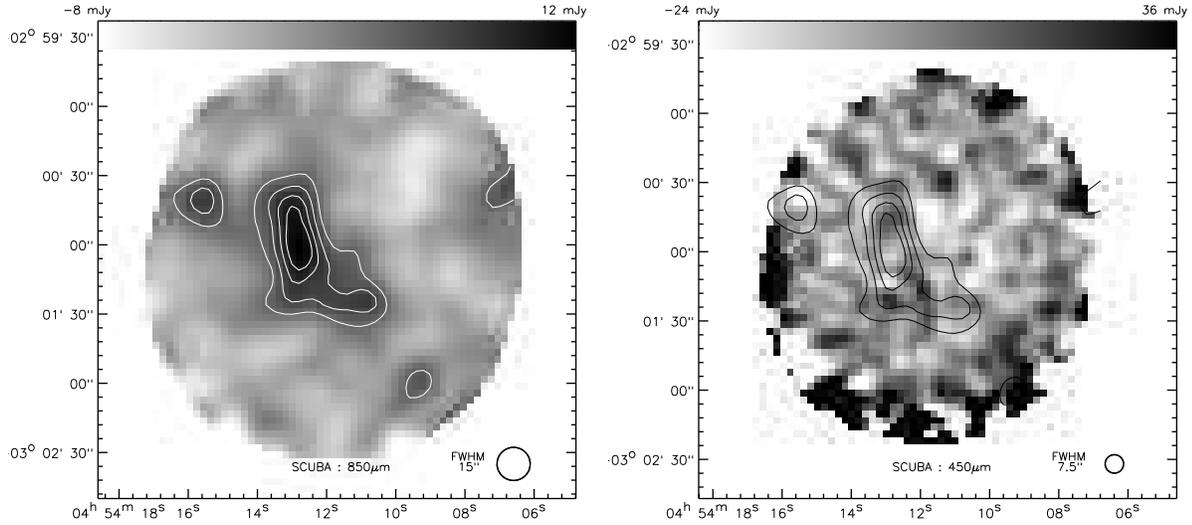}
\caption{SCUBA maps of \ms. The 
850\mum\ flux contours at 4, 6, 8, and 10\,mJy are overlaid on
each. The approximate FWHM of the beam-size at each wavelength is
shown as the circle in the bottom right corner of each map.  The 
850\mum\ image clearly shows emission extended over several beams,
while the short wavelength image shows no discernible flux.}
\label{fig:submmms0451}
\end{figure*}

Because the 1999 data are roughly three times more sensitive, the
gains in co-adding the 1998 data are modest.  Nevertheless the maps
from the 2 separate runs are consistent with each other after the
deconvolution is performed.  We present our best 850\mum\ and 450\mum\
maps in Fig.~\ref{fig:submmms0451}.
The 850\mum\ image shows a continuous ridge of sub-mm
emission, which was partly removed in the chopped data presented in
C02. More importantly, the act of chopping in-field shifted the
centroid south-westward by removing flux on the edge of the arc region.  
The
450\mum\ image, on the other hand, contains no significantly detected
sources.  The lack of a detection of the primary arc with the 450\mum\
array ($3\sigma$ upper limit of 48\,mJy) places a lower limit of
$z\,{\simeq}\,2$ for the source redshift, for reasonable dusty galaxy SEDs.
We also note that since the 1999 data were taken with a 16-point jiggle
pattern (thus fully sampling at 850\mum, but not at 450\mum), the
resulting 450\mum\ map has quite inhomogeneous noise.

\subsection{\hst\ Observations}
\ms\ was imaged with the F702W filter on WFPC2 in 1995 as part of a
program to study luminous X-ray clusters (project ID 5987,
PI: Donahue).  We down-loaded the processed, calibrated image directly
from the archive at the Canadian Astronomy Data Centre.
More recently, the Advanced Camera for
Surveys (ACS; Ford et al. 1998) was used to obtain deep images with
the F775W and F850LP filters (project ID 9292, PI: Ford).  We obtained
the raw data from the public archive and reduced them using the {\sc
multidrizzle} package (Koekemoer et al. 2002).

The WFPC2 F702W images are composed of 4 independent dithered
exposures, with a total exposure time of $10.4\,$ksec.  In each of the two
ACS bands, two exposures totaling $2.4\,$ksec were obtained.  However,
because these observations were not dithered, the resulting ACS images are
prone to detector artifacts and cosmic ray contamination.  This makes
estimating fluxes for some objects in the field problematic.  Astrometry
for the WFPC and ACS images was fixed using 13 objects from the GSC2.2
catalogue. The $5\sigma$ limiting magnitudes for the F702W,
F775W, and F850LP images are 26.3, 24.9 and 25.1 AB magnitudes,
respectively.

\begin{figure*}
\includegraphics[width=\textwidth,angle=0]{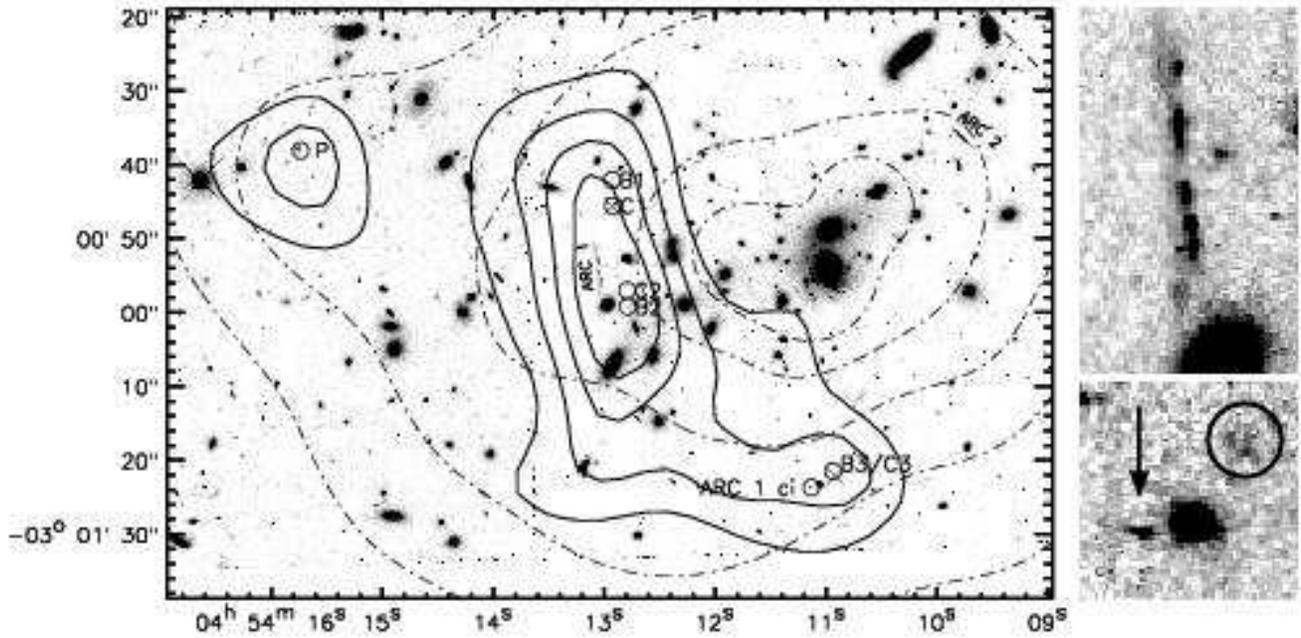}
\caption{\hst\ F702W image of \ms.  {\bf LEFT:} This deep, high 
resolution image
clearly shows the 2 gravitationally lensed arcs previously identified
in lower resolution images (L99).  The stronger arc to the east (Arc
1) appears to be made up us a series of `knots'.  Its counter-image,
`ARC1 ci' is also circled.  Overlaid are the
SCUBA 850\mum\ flux contours with the same levels as in
Fig.~\ref{fig:submmms0451}. The
dotted contours are taken from the medium-band ($1.5$--$4.5\,$keV)
\chandra\ image, with levels at 0.03, 0.05, 0.1, 0.2 and 0.3 photons
cm$^{-2}$ pixel$^{-1}$ (pixel size is 0.492\arcs\ in the \chandra\
image).  Positions of some of the
red objects detected in the NIR images are also
overlaid using circles. X-ray point sources from Molnar et al.~(2002)
coincide with two of these objects, and are shown as crosses. We
also circle the blue point source `P', which is coincident with
the eastern sub-mm point source. {\bf TOP RIGHT:} A close up of 
ARC1 from the F702W image.  {\bf BOTTOM RIGHT:} A close up of 
the ARC1 counter-image (denoted by the arrow).  The circle denotes the
position of B3/C3.  Both close-up images are roughly 4\arcs\ wide.
}
\label{fig:hstms0451}
\end{figure*}

The \hst\ F702W image is presented in Fig.~\ref{fig:hstms0451},
and shows in detail the
lensed arcs reported originally in L99.  ARC1 appears to
contain knots surrounded by diffuse emission which were not previously
resolved from ground-based imaging. These knots display a mirror  
symmetry,
arguing that ARC1 corresponds to a fold-arc (two images merging  
together).
We use this fact later in creating a lensing model for the cluster.
North-east of the cluster centre is a second arc: ARC2, showing a very
smooth light distribution.  VLT spectra described later show that  
ARC1/2 are
at very different redshifts, and thus are not two images of the same  
object.
Furthermore, the colours of these two arcs are quite different
\mbox{($m_{F702W}-m_{F850LP} = 0.3$\,AB} for ARC1 versus 0.6 for  ARC2),
a finding first reported in L99 using lower resolution $V$- and
$R$-band images.

Another notable object in the field is a very blue 
\mbox{($m_{F702W}-m_{F850LP} = 0.0$\,AB)}, morphologically disturbed source (which
we label `P') that seems to be coincident with the SCUBA point source
eastward of the arc. We note also that the \chandra\ contours extend
around this source, suggesting the presence of another massive
component to the cluster.

\subsection{Near IR Observations and EROs}
The CFHT-IR camera on the 3.6-m Canada-France-Hawaii Telescope (CFHT)
was used to image \ms\ in each of the $J$-, $H$-, and \kpd band filters.
The images were obtained in moderate conditions, with seeing roughly
1.0 arcsec. In each filter, observations were dithered in a hexagonal
pattern, providing uniform coverage of a $3\times3$ arcminute region.
Several short ($<90\,$second) integrations were taken, for a total
exposure time of $7.2\,$ksec in $H$, $5.6\,$ksec in \kp, and $6.8\,$ksec in
$J$.  Observations of the standard FS11 were taken to measure the zero
points. The limiting magnitudes in the $J$-, $H$- and \kpd band
images were 24.4, 23.5, and 23.8 (AB, $5\sigma$) respectively.

Data were reduced using the {\sc dimsum} package in IRAF, together with
{\sc sextractor} to detect objects and estimate fluxes within 2
arcsecond apertures. We compared the \kpd selected
catalogue against 84 objects
detected in the shallower survey of Stanford et al.~(1998). Although
there is no evidence of bias between our catalogues, we did find an
RMS scatter of 0.1 magnitudes between sources in each of the three
NIR bands, and we
use this as the lower limit to the 
error estimate for our $J$-, $H$-, and \kpd band magnitudes.  Our magnitudes
also agree with a deeper Subaru NIR image described in Tanaka et al. (2003).

\begin{figure}
\includegraphics[width=0.5\textwidth,angle=0]{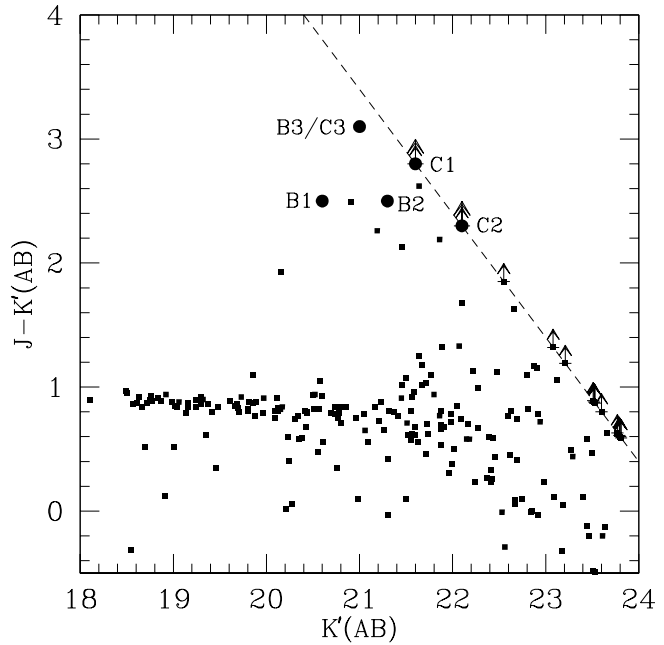}
\caption{Colour-magnitude diagram of \ms\ using the NIR data. The
region above the dashed line is beyond the limit of the survey.  Only
objects within 1\arcm\ of the BCG are plotted.  Labeled sources
denote the triply-imaged ERO pair described in the text. None of the
points have been corrected for lensing amplification.  All fluxes
were measured in 2\arcs\ apertures, except for B2, which required a
1\arcs\ aperture due to its proximity to nearby sources in the image.}
\label{fig:cmd}
\end{figure}

\begin{figure*}
\includegraphics[width=1.0\textwidth,angle=0]{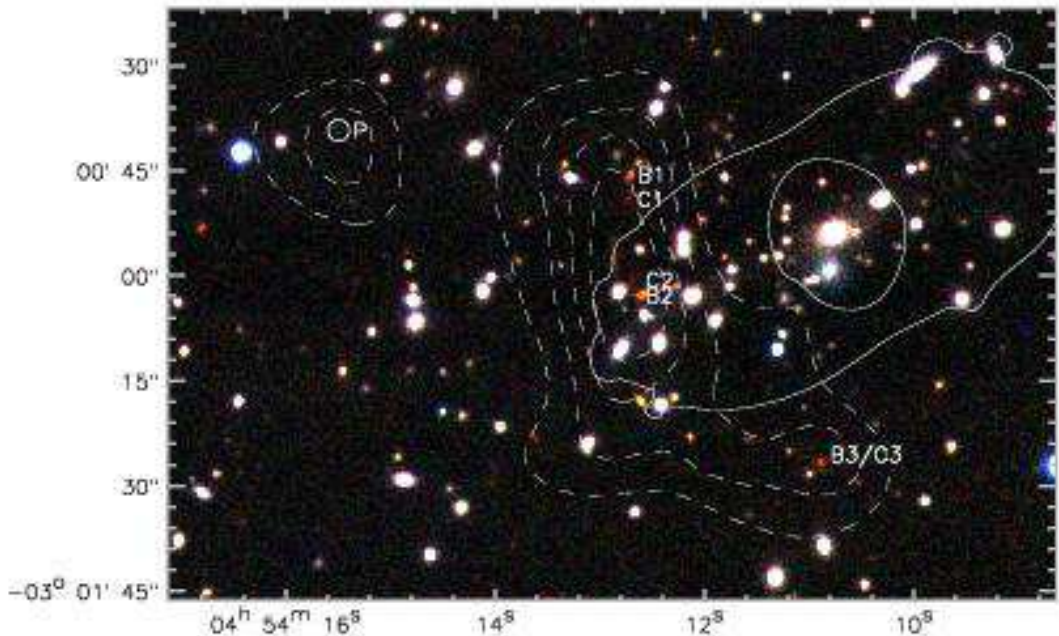}
\caption{A true-colour image of \ms\ using the NIR data.  The scale is
approximately the same as in Fig.~\ref{fig:hstms0451}.
The three images of the ERO pair are labeled here,
as well as the blue source `P' (although it is essentially invisible in the  
NIR). The outer white contour traces shows the $z=2.911$ critical 
tangential line, and the inner one corresponds to the radial line.
Sub-mm contours at the same levels in Fig. 2 are shown as the dashed lines.}
\label{fig:nirms0451}
\end{figure*}

A colour-magnitude diagram (Fig.~\ref{fig:cmd}) using the $J$ and
\kp\ magnitudes, and a false colour image of the cluster created
using the NIR images (Fig.~\ref{fig:nirms0451}) reveals objects with  
extremely
red colours $(J-K^\prime>2.3)$.   In Section \ref{sec:lensing} we will
describe the lensing model
of the cluster, but it is
important here to highlight sources B and C, which are a pair of EROs
re-imaged in 3 places due to strong lensing.  The first and second
image, sources B1/C1 and B2/C2 are mirrored across the critical line
in the lens model, and have similar magnification factors.  Unfortunately
B2/C2 are in a particularly crowded region of the image, making accurate
photometry difficult.
At the
position of B3/C3, the lensing is much weaker, and thus the object
separation decreases to the point where the two distinct objects
overlap.

The connection between sub-mm galaxies and \textsl{some} EROs is well
established; dust responsible for the sub-mm emission also reddens the
optical-UV spectrum.   NIR bright objects invisible in deep \hst\
imaging were found coincident with sub-mm sources in observations
towards other lensing clusters (Smail et al. 1999). 
Roughly one-third of the sub-mm galaxies
detected in the UK 8-mJy survey have an ERO as the most likely
counterpart (Ivison et al. 2002).  In addition, Webb et al.~(2004) claim
that pairs of EROs denote regions of sub-mm emission, though not
necessarily the source of it.  And in general,
stacking analyses of optical and
NIR selected galaxies in sub-mm fields show a convincing trend for
redder objects to be statistically more SCUBA-bright (Wehner et
al. 2003, Borys et al 2004).  These findings, along with the fact
that the 3 ERO pairs here trace out the sub-mm contours quite
convincingly, make this a strong candidate for the sub-mm arc
counterpart.

\subsection{\chandra\ observations}
\ms\ was observed with the \chandra\ Advanced CCD Imaging Spectrometer
(ACIS) detector for a total of $41\,$ksec. Full processing details are
contained in D03. The astrometry for a \chandra\ ACIS-S observation is
typically accurate to about 1.0\arcs, but can have errors as large as
3.0\arcs.  X-ray point source fluxes are described in Molnar et
al.~(2002). Two sources are detected within the field of the SCUBA
observations, and are coincident with C1 and C3.  The X-ray source
coincident with C1 was detected with $45.2\pm8.1$ counts, while that
of C3 has $21.4\pm5.7$.  The ratio of counts is the same as the ratio
of magnification factors derived from the lensing model described in
the next section.

Since a point source was not reported in D03 at the position of C2,
we estimated the X-ray flux using a 2\arcs\ aperture.  The lensing model
predicts that that C1 and C2 have approximately
the same amplification factor, thus we expect $\sim45$ counts.
However, we only find $13\pm6$ counts
between 0.5--$2.0\,$keV at the position of C2.
We note that photons from C2 pass through more
of the cluster gas, since it is closer to the brighter X-ray centre
than C1 (see Fig.~\ref{fig:hstms0451}).
If we assume that the NIR fluxes are accurate,
then the fainter C2 (by 0.5 magnitudes in \kp) suggests the number
of counts from C2 should be around 25.  Our estimate  
of 13 counts is still low compared to this expectation, but given the  
difficulty in extracting an X-ray flux estimate of point
sources on top of a brighter, diffuse background (the cluster gas), it does
not seem unreasonable.  More accurate NIR imagery, coupled with more  
detailed modeling of extinction through the cluster, could resolve this  
issue.  Based on the strength of the lensing model and colours/geometry
of the EROs, we still conclude that the three B/C pairs are the same
objects, with C also being an X-ray source (despite the apparent
non-detection of X-rays in C2).

The X-ray profile discussed in D03 allows us to predict the SZ
increment at 850\mum, which, in the absence of chopping, peaks at
roughly $3\,$mJy. By applying the astrometry from the sub-mm data, we
constructed a simulated map of what the SZ signal would look like when
the chopping is accounted for. The peak drops to $2\,$mJy and the
SZ emission falls
off to roughly $1\,$mJy at the position of the main arc.  If we
subtract this shape from the data it makes little difference to the
flux levels or SNRs in the SCUBA map.  On the other hand, the
contamination of so much lensed sub-mm emission makes this particular
field useless for sub-mm SZ studies, at least until much higher
resolution images can be obtained (Zemcov et al. 2003). Indeed,
measuring the SZ increment in {\em any} strong lensing cluster will 
prove difficult simply because the resolution is too poor to
remove the enhanced number of point sources detected in the field.

\subsection{VLT spectrum of ARC1}
Deep multi-object spectroscopy of \ms\ was obtained in service mode
for the FORS2-GTO programme: 68.A-0015 (PI: Appenzeller) on December
10th, 11th, 12th and 21st 2001, with the FORS-2 instrument
(Appenzeller \& Rupprecht 1992) on the VLT-Yepun (UT4). We retrieved
these data from the ESO/VLT archive and reduced them using standard
{\sc IRAF} procedures. Two masks were prepared (`M128' and `M129') and
exposed with the high throughput/low resolution grism GRIS\_150I
(5.2\AA/pixel). For each mask five exposures of 2.9\,ksec each were
obtained (totaling 14.5\,ksec).  ARC1 is part of the `M129' mask,
and ARC2 of the 'M128' mask.
The four nights had variable seeing, ranging from 0.6 to 1.0$\arcsec$,
and were likely not fully photometric.  Nevertheless we obtained a
crude flux calibration of these observations using the
spectrophotometric standard star GD108.

\begin{figure*}
\includegraphics[width=0.4\textwidth,angle=270]{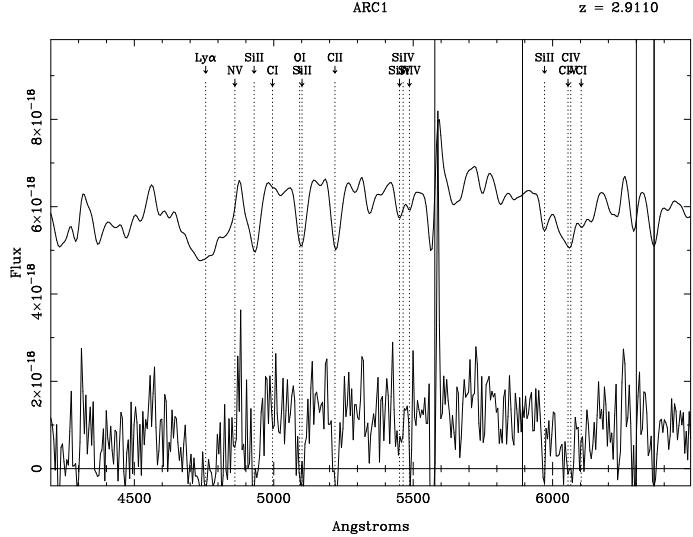}
\caption{FORS2 spectrum of ARC1 plotted in the observers rest
frame.  The bottom curve is the raw spectrum, and a smoothed version
is plotted above it.  Based on the strong absorption features, the
redshift of the arc is constrained to be $z=2.911\pm0.003$.
Ly$\,\alpha$ absorption is evident near 4766\AA, which is seen in
only about 20\% of the submillimetre galaxy spectra in the sample
of Chapman et al.~(2004).}
\label{fig:spectra}
\end{figure*}

ARC2 displays only one strong emission line at $\lambda=7144$\AA. We
identify it as O{\sc ii}, and thus calculate a redshift of  
$z=0.917\pm0.001$.
The derived redshift for ARC1  (Fig.~\ref{fig:spectra}) is
$2.911\pm 0.003$ based on the strongest interstellar absorption lines:
Si{\sc ii}$\,\lambda$1260.42\AA, O{\sc i}$\,\lambda$1302.17\AA\ and
C{\sc ii}$\,\lambda$1334.5\AA\ as well as C{\sc iv}$\,\lambda$1548.19 and
1550.77 in absorption. There is also a hint of the triple Si{\sc vi}
lines at $\sim\lambda$1397\AA\ and the Si{\sc ii}$\,\lambda$1526.70\AA\  
line.

The ARC1 spectrum lacks AGN signatures sometimes found in the spectra of
radio selected sub-mm galaxies.  Moreover, it appears to show Ly$\,\alpha$
in absorption, a feature seen in only about 20\% of
the spectroscopic survey of 
$\sim100$ sub-mm galaxies conducted by Chapman et al.~(2004).
That ARC1 does not look similar to most known
sub-mm galaxy spectra does not necessarily mean it is not the host of
the sub-mm emission.  The set of blank-field sub-mm galaxies with
spectroscopically determined redshifts are biased towards galaxies
with strong Ly$\,\alpha$ emission, and
it is clear we are studying a sub-mm galaxy which is intrinsically
much fainter (and possibly
much different) than those reported in Chapman et al.~(2003, 2004).

\section{A lensing model for \ms}
\label{sec:lensing}
In our first attempt to model the system we employed a simple
isothermal ellipsoid for the cluster mass (using parameters from D03),
and a second elliptical mass distribution to model the Brightest Cluster
Galaxy (BCG).
This allowed us to gauge the general properties of the lens,
and showed that the ridge of sub-mm emission generically lies along the
critical line for high redshift sources.  However, the
importance of determining the precise position and magnifications of
the multiply lensed EROs necessitated a more detailed approach.
Indeed, as the \chandra\ contours
in Fig.~\ref{fig:hstms0451} show, the mass within the central 1\arcm\  
of the
cluster is not smoothly distributed, and the BCG position and X-ray
centroid are not coincident.

To model the mass distribution of this cluster we used both a cluster
mass scale component and cluster galaxy mass component in a similar
way to Kneib et al.~(1996; see also Smith et al.~2003).  Using the
\textsc{lenstool} software (Kneib~1993) we interactively implemented the
lensing constraints.  The high resolution \hst\ data show
a clear mirror symmetry along the ARC1, making it the
merging of 2 images (fold arc). A simple
elliptical model allows us to identify its counter-image 3\arcs\
to the south-east of B3/C3,
which is much less amplified (see Table~\ref{tab:objects}).

Similarly, the ERO B/C pair can then be included in the lensing model.
B1/C1 and B2/C2 are straddling the critical line at redshift $z=2.9$,
thus by placing the B/C pair at $z\sim2.9$, we easily identify B3/C3
as the third image of the B/C pair.

By combining the ARC1 and B/C constraints, we refined the mass model,
and are able to place more precise limits of \mbox{$z=2.85\pm0.10$} for  
the ERO B/C pair. It is therefore likely that both ARC1 and B/C are at
the same redshift. If it is the case, then the separation between
the LBG and the ERO pair in the source plane is only $\sim10$kpc, suggesting
that these 3 galaxies are interacting.

Using the photometry derived from the optical and  near-infrared
data, as well as the magnifications estimated from the lensing model, we
summarize, in Table~\ref{tab:objects}, the list of objects highlighted
in Fig.~\ref{fig:hstms0451} and Fig.~\ref{fig:nirms0451}.

\begin{table*}
\caption{Photometry for objects of interest in the field of \ms.  These
limits are shown as $5\sigma$, and all are reported in the AB
magnitude system. None of the fluxes are corrected for lensing magnification.
All magnitudes are measured in 2\arcs\ apertures, except for the
arcs (measured using ellipses covering the same area in each of the
fields, with the counter-image ARC1ci measured using a 1\arcs\ aperture because
of possible contamination from a nearby source) and B2, where a
magnitudes in a 1\arcs\ aperture is also given because of possible flux
contamination. 
The final column gives magnifications from the lensing model; Most are well 
constrained, except for ARC1.  Since it lies right along the critical line,
an more accurate estimate of the amplification is difficult.
}
\begin{tabular}{lllllllllll}
\hline
ID        & RA          & DEC          & F702W        & F775W        &  F850LP       & $J$            & $H$          & \kp\ & $J-$\kp  & Mag. \\\hline
ARC1      & 04:54:13.0  & $-$03:00:55.9  & $22.9\pm0.1$ & $22.6\pm0.1$ & $22.6\pm0.1$ & $21.2\pm0.1$ & $21.0\pm0.1$ & $20.2\pm0.1$ & 1.0    & 50--100     \\
ci(1\arcs)& 04:54:11.1  & $-$03:01:26.6  & $25.7\pm0.4$ & $25.2\pm0.2$ & $25.3\pm0.3$ & $>24.4$      & $>23.5$      & $>23.8$      & \dots  & 4.7$\pm0.5$ \\
B1        & 04:54:12.8  & $-$03:00:44.1  & $24.9\pm0.3$ & $25.0\pm0.2$ & $24.4\pm0.2$ & $23.1\pm0.1$ & $21.4\pm0.1$ & $20.6\pm0.1$ & 2.5    & 8$\pm1$     \\
B2        & 04:54:12.7  & $-$03:01:01.3  & $>26.3$      & $>24.9$      & $>25.1$      & $22.9\pm0.1$ & $22.2\pm0.1$ & $20.7\pm0.1$ & 2.2    & 10$\pm1$    \\
B2(1\arcs)&             &                & $>26.3$      & $>24.9$      & $>25.1$      & $23.8\pm0.2$ & $22.9\pm0.2$ & $21.3\pm0.1$ & 2.5    &             \\
C1        & 04:54:12.8  & $-$03:00:47.6  & $>26.3$      & $>24.9$      & $>25.1$      & $>24.4$      & $>23.5$      & $21.6\pm0.1$ & $>2.8$ & 10$\pm1$    \\
C2        & 04:54:12.7  & $-$03:00:59.1  & $>26.3$      & $>24.9$      & $>25.1$      & $>24.4$      & $>23.5$      & $22.1\pm0.1$ & $>2.3$ & 11$\pm1$    \\
B3/C3     & 04:54:10.9  & $-$03:01:24.5  & $25.2\pm0.2$ & $>24.9$      & $>25.1$      & $24.1\pm0.2$ & $21.8\pm0.1$ & $21.0\pm0.1$ & 3.1    & 5$\pm1$     \\
P         & 04:54:15.6  & $-$03:00:38.7  & $24.0\pm0.2$ & $23.9\pm0.1$ & $24.0\pm0.2$ & $>24.4$      & $>23.5$      & $>23.8$      & \dots  & 1.5         \\
\end{tabular}
\label{tab:objects}
\end{table*}

\section{Discussion}
With the lensing model in hand, we are now able to further examine  
which galaxy hosts the sub-mm emission.
Apart from the (we believe remote) possibility that the sub-mm emission
arises from other sources entirely, there are 3 obvious possibilities,
which we discuss in turn.  We also comment on the additional sub-mm source
to the east of the arc, which we label `P'.

\subsection{Scenario 1: The blue arc as the source of the sub-mm  
emission.}
Although the spectrum for ARC1 does not match that of a `typical' sub-mm
galaxy from the Chapman et al.~(2004) sample, it is similar to the most
absorbed quartile of the Lyman break galaxies (LBGs), which generally  
correlate
with the reddest, most dust extincted varieties (see Shapley et
al. 2003).  It is this dust which can be responsible for the sub-mm
emission, though using the rest frame UV spectra of LBGs to estimate
the sub-mm flux is notoriously difficult (Chapman et al. 2000).  In a
sample of $\sim30$ Lyman break galaxies with the most extreme
reddening (and hence most likely chance to be sub-mm bright), only one
(MMD-11) was detected.

Indeed, several other studies find a similar lack of sub-millimeter
emission from individual LBGs (Webb et al. 2003, Borys et al. 2004).
On the other hand, Peacock et al.~(2001) stack LBGs in the deep
sub-mm map of the Hubble Deep Field (Hughes et al. 1998)
to estimate that the {\em average\/} 850\mum\ flux from an LBG is
\mbox{$0.20\pm0.04\,$mJy/M$_\odot h^{-2}$yr$^{-1}$}.  Given that the  
unmagnified sub-mm flux would be $0.1-0.2\,$mJy if the LBG is 
the host, then it is entirely
possible that we have simply detected an ordinary LBG producing stars
at a rate of a few solar masses per year.

An argument against ARC1 being the sole sub-mm source 
is the fact that the sub-mm emission is much
more extended than the optically detected arc.  It extends at least 15
arcseconds further north (a full SCUBA beam-width).  Over 45
arcseconds (3 beams) separate the arc from B3/C3, where the toe of the
sub-mm emission to the south-west tapers off.  Note that the ARC1
counter-image is too faint to be responsible for the toe, since the
amplification factors predicted from the lens model are not consistent  
with the observed sub-mm flux.  Even
assuming the lowest reasonable magnification factor for ARC1,
we would predict only $\sim1\,$mJy of 850\mum\ flux at the position of
the ARC1 counter-image.

\subsection{Scenario 2: Contribution of the multiply lensed EROs to the  
sub-mm flux.}
Galaxies with very red colours often suggest the presence of large
amounts of dust, and several sub-mm blank field surveys (Ivison et
al. 2002, Webb et al. 2003, Borys et al. 2004) have found such objects
coincident with sub-mm detections.  The triply-imaged ERO pair in
\ms\ is therefore a reasonable candidate for the counterpart to the
sub-mm arc.  The relative fluxes of objects B1/C1 and B3/C3 match the  
relative
amplifications predicted from the lensing model.  B2 is a little
difficult to  compare,
given the uncertainty in the flux due to the contamination from a  
nearby source,
and the discrepancy with C2 has already been discussed.

Unlike the last scenario, the position of the three
images can easily accommodate the extremities of the extended emission.
The sub-mm contours around B3/C3 suggest that it is a SCUBA point
source, though the signal-to-noise is insufficient to be certain.  
Assuming B/C
is the host of the sub-mm emission, and using the peak 850\mum\
flux of 12\,mJy, we would predict 6\,mJy at the position of B3/C3.  This
is essentially what is observed.  If the EROs are the host of the
sub-mm emission, then the unmagnified flux is $\sim1.0$\,mJy at
850\mum.

It is also worth noting the several red objects in southern part of field.
These are in the same region as the southernmost extension of the sub-mm arc,
and might be contributing sub-mm flux:  Using colour-colour plots and the 
Pozzetti \& Mannucci (2000) criteria for distinguishing dusty starbursts from
 passively evolved ones, Tanaka et al. (2003) show that both object B and an 
red object 4\arcs\ to the northeast of B3/C3 are consistent with dusty 
starbursts. If so, this would help explain the extended sub-mm emission 
between B3/C3 and its counter-images.

\begin{figure}
\includegraphics[width=0.5\textwidth,angle=0]{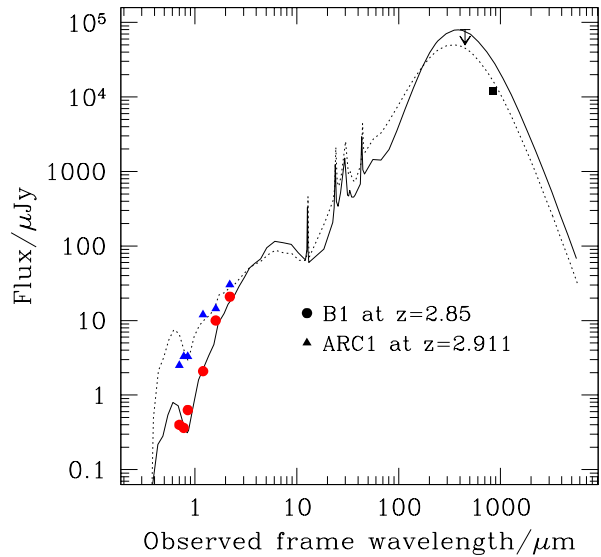}
\caption{Spectral Energy Distribution deduced from the \hst\ and NIR data
for the ERO B1 (circles) and ARC1 (triangles). The square point is the
850\mum\ emission of sub-mm arc.
The 5$\sigma$ upper limits is also shown for the 450\mum\ measurement.
The solid line is a model template redshifted to $z=2.85$, and scaled to
match the NIR measurements of B1.  The dotted line is another model  
redshifted
to $z=2.911$ and scaled to match the NIR data from ARC1.
These models are derived from Efstathiou,
Rowan-Robinson, \& Siebenmorgen (2000), and assume a 72\,Myr old
star-burst in the case of B1, and 45\,Myr for ARC1.
Though the NIR fluxes are close fits to the model by
design, the measured 850\mum\ flux is only a factor of 2 from the
predictions.}
\label{fig:seds}
\end{figure}

In Fig.~\ref{fig:seds} we plot the optical fluxes for the best detected  
object
from this trio of pairs, B1, on a template SED obtained from the
library of Efstathiou, Rowan-Robinson, \& Siebenmorgen (2000).  These
combine the Bruzual \& Charlot stellar population synthesis models 
(Bruzual \& Charlot 1993)
with a radiative transfer code , and produce SED templates that match
local star-burst galaxy spectra.  We translated the templates to
$z=2.85$ and scaled them to match the NIR fluxes.  As  
Fig.~\ref{fig:seds}
shows, there is reasonable agreement between the
assumed redshift, the measured photometry, and the SED template.
There is some freedom in these models, and hence we have not attempted
to provide
detailed constraints on the star-burst age, formation history, and
composition.  The main point is that the data support the claim that the
EROs can be responsible for the sub-mm flux. Ground based NIR  
spectroscopy,
as well as \spitzer\ observations, will be of considerable help in  
determining the nature of this galaxy pair.

\subsection{Scenario 3: Contribution from both}
\begin{figure}
\includegraphics[bb=0 0 440 440,width=0.4\textwidth,angle=0]{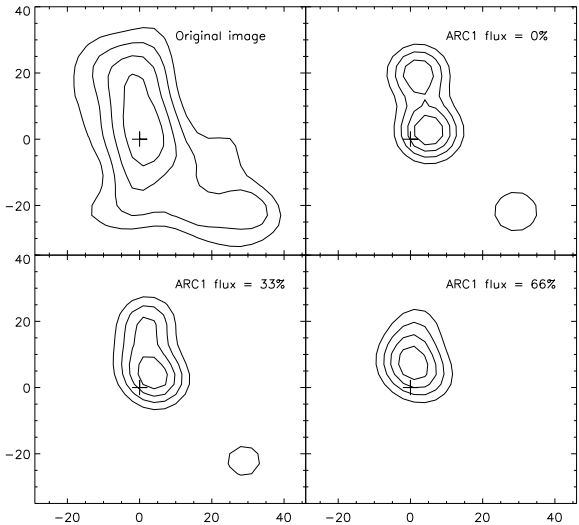}
\caption{Simulating the relative contributions of the LBG and ERO.
The plus symbol in each panel denotes the centre of ARC1,
and the axes are in units of arcseconds. In the
top left panel we show the 850\mum\ contours at 4, 6, 8 and 10\,mJy from
the original sub-mm image. The top right shows a simulation where a
source was placed at the position of B1, B2, and B3 (with relative
fluxes determined from the lensing model), but no flux
from ARC1.  The bottom left shows a model
where the LBG arc contributes 4\,mJy, and then 8\,mJy in the last panel.
These simulations suggest that that at least some flux is necessary from
{\it both\/} ARC1 and B/C in order
to reproduce the observed sub-mm image. }
\label{fig:sims}
\end{figure}

We have now established that both objects are capable of producing
sub-mm flux, so naturally we need to ask if both objects are
contributing to the sub-mm arc.  To test this, we created a blank map
the same size as the SCUBA data, and inserted sources at positions
corresponding to B1, B2 and B3, as well as ARC1.  The
relative fluxes between B1, B2 and B3 was fixed by the lensing model
predictions for amplification, leaving the flux ratio between ARC1 and
B1 as the only remaining parameter.  The final constraint is that the
peak flux needs to match that observed, which in this case is 12\,mJy.
The results of these simulations are shown in Fig.~\ref{fig:sims}.

We found that no contribution from ARC1 resulted in B1/2 being
resolvable, which does not match the observed sub-mm image.
In addition,
ARC1 needs to contribute at least 2\,mJy in order to `fill in'
the gap between B1/2.  At more than 7\,mJy, ARC1 dominates the contours,
and fails to reproduce the extended nature of the source. Although
it is difficult to precisely constrain the relative
contributions, this exercises suggests that sub-mm emission from both
the LBG arc and the ERO pair are required to explain the
morphology and brightness of the sub-mm arc, with $\sim2/3$ of the observed
flux coming from the ERO B/C pair.

\subsection{The sub-mm point source, `P'}
Though the ERO pair and LBG arc just discussed are the focus of this
paper, it is worth noting the blue galaxy `P', which seems coincident  
with the
SCUBA detected point source north-east of the arc.  Because there is
simply no other optically detected galaxy within the error circle
(6\arcs) of the SCUBA source, it is reasonable to conclude that the
objects are at least associated with each other, even if they are not
necessarily the same thing.

Note in Fig.~\ref{fig:hstms0451} how the X-ray contours bend around
this object.  This suggests the presence of another massive component
interacting with the cluster, or perhaps a projection effect from another
X-ray cluster.  If the sub-mm object is behind the cluster, then
it is only weakly lensed; the prediction is $\sim1.5$ from our lensing
model, meaning that the
unmagnified 850\mum\ flux for this object would be about 4\,mJy.
The lack of a 450\mum\ detection is not very constraining.  Using a wide  
range of temperatures and emissivities in a single dust component SED, we  
predict a 450\mum\ flux ${<}\,48$ for $z>0.5$.  Since the 450\mum\ map  
has a $3\sigma$ upper limit comparable to this, we cannot rule out that
the sub-mm source is actually coming from the cluster itself.  This would be
an interesting case, since it would suggest that star-formation is occurring
in a massive sub-component merging with \ms.  Further spectroscopic and
X-ray studies may unravel this particular mystery.

\section{Conclusions and future work}
We have presented broad-band sub-mm and near-IR images of the cluster
\ms\ and attempted to understand the nature of the highly extended
sub-mm emission.  The only other case of a multiply lensed sub-mm 
detection in a cluster field  has recently been reported in Abell 2218 
(Kneib et al.~2004).  They identified the source of the triple sub-mm 
image as a LBG galaxy (showing a complex structure, indicative
of a close merger) in a $z\sim2.5$ group.  

In the case of \ms, the evidence suggests that the sub-mm emission is coming 
from at least two components:
a blue arc at $z=2.911$ and a multiply imaged ERO pair predicted from
the lens model to be at $z\sim2.85$. 
Using a simple model for the 2 sources, we estimate that the multiply 
lensed EROs contribute $\sim$2/3 of the observed flux in the sub-mm arc. 
Correcting for the lensing amplification, we estimate that the unlensed
fluxes at 850\mum\ are 0.1\,mJy for the LBG and 0.4\,mJy for the ERO pair.

The uncertainty in the redshift for the ERO keeps the possibility open
that the ERO pair and LBG are at identical redshifts.  This would place them
only $\sim$10 kpc apart at $z\sim2.9$, and implies that the objects are interacting
and their interaction is likely to be the origin of a violent star-burst
revealed by the strong sub-mm emission.  To confirm that the LBG is 
effectively spatially connected with the ERO pair can only be determined 
by measuring the redshift of the EROs using NIR spectroscopy or via
millimeter CO line search; Given the constraints on the redshifts of 
either candidate,
sub-mm CO observations can be used to detect the CO(3-2) transition line
that falls within the observation window of several ground-based
heterodyne receivers.

To verify these conclusions, and to accurately separate the 
contributions from each component will
require sub-mm/mm interferometry.  The SCUBA flux is bright enough
that it should be detectable using the IRAM Plateau de Bure
interferometer.  The SMA and ultimately ALMA will also have the power
to resolve the components in great detail.
With such further study, and using the lensing
amplification of nature's telescope, we should be able to provide a full
accounting of the source of sub-mm emission in this intriguing system.

\section*{Acknowledgments}
CB thanks the referee, Ian Smail, for comments that greatly improved
the paper, Margrethe Wold and Duncan Farrah for useful discussions
regarding NIR data reduction, as well as Dan Durand and Luc Simard at
HIA for their assistance in recovering the ACS data. We also thank
Stella Seitz for discussions regarding the VLT spectra, as well as
Linda Tacconi and Dieter Lutz.
This work was
supported in part by the Natural Sciences and Engineering Research
Council of Canada. JPK acknowledges support from Caltech and CNRS.
The James Clerk Maxwell Telescope is operated by The Joint Astronomy  
Center on
behalf of the Particle Physics and Astronomy Research Council of the
United Kingdom, the Netherlands Organisation for Scientific Research,
and the National Research Council of Canada. The \hst\ data were
recovered from the Canadian Astronomy Data Centre (CADC), which is
operated by the Herzberg Institute of Astrophysics, National Research
Council of Canada. The Guide Star Catalogue II is a joint project of
the Space Telescope Science Institute and the Osservatorio Astronomico
di Torino.

\appendix
\section{Map-making by direct matrix inversion}
To make maps from data taken with SCUBA, one normally adds the
measured signal into the pixel corresponding with the position of a
bolometer in the `on' position. For observations with in-field
chopping however, this will result in a map with negative echos.
Since the \ms\ field appears to be quite dense in bright sub-mm
sources, it is necessary to deconvolve the chop to obtain a
single-beam map.  This can be posed as a linear algebra problem.
Following the notation given in Stompor et al.~(2002), the $M$
measurements that make up the time-stream data vector, $\bf{d_{\textsl
t}}$ are related to a pixelized region of sky, $\bf{m_{\textsl p}}$
that has $N$ elements via a sparse M$\times$N mapping matrix,
$\cal{A}$:
\begin{equation}
\label{equ:map}
\bf{d_t}=\cal{A}\bf{m_p} + \bf{n_t}
\end{equation}
The vector $\bf{n_t}$ is the noise estimate for each  
measurement with statistical properties described by
$\bf{{\cal N}=<n_tn_{t^\prime}>}$. The best estimate of  
the flux in the pixelized map, $\bf{m_{\textsl p}}$, in a least squares  
sense, has a well known solution (Stompor et al. 2002, Tegmark 1997,  
Wright 1996):

\begin{eqnarray}
\label{equ:minvarmap}
\bf{m_{\textsl p}}&=&\left({\cal AN}^{-1}{\cal A}^T\right)^{-1}\,{\cal AN}^{-1}\bf{d_t},\\
\label{equ:minvarnoise}
{\cal M}&=&\left({\cal AN}^{-1}{\cal A}^T\right)^{-1}.
\end{eqnarray}

${\cal M}$ is an $N\times N$ matrix that measures how much the 
pixels are correlated with each other. ${\cal M}_{ii}^{1/2}$ 
is the formal error estimate for pixel $i$.
${\cal A}$ is typically extremely sparse, with each row here
having only three non-zero entries [$1.0, -0.5, -0.5$] corresponding
to the pixels that the primary and 2 off bolometers hit.  Given that
SCUBA observations are differential in nature, the matrix
${\cal AN}^{-1}{\cal A}$ is formally singular, since
there exists a non-zero vector $\bf{m_p}$ where 
${\cal A}\bf{m_{\textsl p}}=0$.  
Specifically, and unsurprisingly, the degeneracy is in the
overall mean of the map, which SCUBA is insensitive to (i.e., setting
$\bf{m_p}$ to a constant).

${\cal N}$ measures how correlated the noise is in the time
domain.  For anything but uncorrelated noise this matrix is difficult
to manage without special techniques, since ${\cal N}$ is a $M\times M$
matrix and $M$ can easily reach $10^6$
samples.  The combined dataset for \ms\ has $1.14/0.52\times10^6$
samples at 450/850\mum\ , corresponding to 12 hours of integration
time.

Assuming uncorrelated noise, the matrix 
${\cal AN}^{-1}{\cal A}$ and vector ${\cal AN}^{-1}\bf{d_{\textsl t}}$ 
can be populated in time order, requiring
${\cal O}(M)$ operations.  Thus we are faced only with the task of
inverting an $N \times N$ matrix.  850\mum\ jiggle maps are typically
binned into 3\arcs\ pixels in order to reasonably sample the beam.
Given the 2.3\arcm\ field of view of a jiggle map observation, the
number of pixels is on the order of a few thousand, which is small
enough that any modern workstation can invert the required matrix and
obtain the solution in an acceptably short time.

The only caveat is in the selection of `dummy pixels', which are used
to account for positions a bolometer may hit that are outside the
solution region.  For example, the 1999 data used 180 arcsec
chop-throws, which place the bolometers in the off position well
outside the central target region.  The procedure we adopt is to
assume that there is no signal outside the target region, and assign a
single dummy pixel. To verify the procedure, simulated data-sets were
created, based on the astrometry information in the \ms\ observations.
False sources were placed such that some of the time the chop would
cancel out their signals.  To correct for the mean of the map (which
is forced to zero), we select a region of the map free of sources and
compute the average signal, which is then applied to the whole map
(there is some subjectivity to this procedure, since for real maps one
will not know for certain which regions are devoid of signal).  The
$\chi^2$ for the reconstructed maps compared against the input source
map was always within $1\sigma$ of the number of pixels in the map.
In doing these tests, we also tested varying the number and position
of dummy pixels to verify that they did not adversely affect the
inversion.

This algorithm was then applied to the actual \ms\ data.  First, maps
from the 2 separate runs were solved and compared.  The $\chi^2$
between these half maps is better than that obtained from the raw
maps. The overall improvement was not very dramatic but this is not
completely unexpected since the chop configuration only causes 2 of
the sources in the map to interfere with each-other.  Also, the 1999
data are of superior quality, and hence is weighted higher in the
inversion. However, the improvement in consistency between the half
maps is dramatic in the regions where there are chopped sources.

Although the gain may be marginal in some cases, in general this
direct inversion technique can be straightforwardly applied to SCUBA
jiggle-map data.It is easy to automate the procedure, with the only
required input being how to treat bad bolometers, how to assign pixels
outside the target region, and what to do about the DC level.

\end{document}